\documentclass[preprint]{aastex}
\usepackage{color}

\shorttitle{Baryons in the DM Halo of NGC5005}
\shortauthors{Richards et al.}

\begin{document}

\title{Baryonic Distributions in the Dark Matter Halo of NGC5005}

\author{Emily E. Richards,$^1$ L. van Zee,$^1$ K. L. Barnes,$^1$ S. Staudaher,$^2$ D. A. Dale,$^2$ 
T. T. Braun,$^1$ D. C. Wavle,$^1$ D. Calzetti,$^3$ J. J. Dalcanton,$^4$ J. S. Bullock,$^5$ and R. Chandar$^6$}
\affil{$^1$Indiana University, 727 East 3rd Street, Swain West 318, Bloomington, IN 47405, USA}
\affil{$^2$University of Wyoming, 1000 E. University, Dept 3905, Laramie, WY 82071, USA}
\affil{$^3$University of Massachusetts, Department of Astronomy, LGRT-B 619E, 710 
North Pleasant Street, Amherst, MA 01003, USA}
\affil{$^4$University of Washington, Box 351580, U.W., Seattle, WA 98195, USA}
\affil{$^5$University of California, Irvine, Department of Physics \&~Astronomy, 
4129 Frederick Reines Hall, Irvine, CA 92697, USA}
\affil{$^6$University of Toledo, 2801 West Bancroft Street, Toledo, OH 43606, USA}
\email{er7@indiana.edu}

\begin{abstract}
We present results from multiwavelength observations of the galaxy NGC5005. We use new
neutral hydrogen (HI) observations from the Very Large Array to examine the neutral gas morphology
and kinematics. We find an HI disc with a well-behaved flat rotation curve in the radial range 
20\arcsec-140\arcsec. Ionized gas observations from the SparsePak integral field unit on the WIYN 
3.5m telescope provide kinematics for the central 70\arcsec. We use both the SparsePak and HI velocity 
fields to derive a rotation curve for NGC5005. 
Deep 3.6$\mu$m observations from the {\it Spitzer Space Telescope} probe the faint extended stellar 
population of NGC5005. The images reveal a large stellar disk with a high surface brightness component 
that transitions to a low surface brightness component at a radius nearly 1.6 times 
farther than the extent of the gas disk detected in HI. The 3.6$\mu$m image is also decomposed into bulge 
and disk components to account for the stellar light distribution. Optical broadband B and R and narrowband 
H$\alpha$ from the WIYN 0.9m telescope complement the 3.6$\mu$m data by providing information about the 
dominant stellar population and current star formation activity. 
The neutral and ionized gas rotation curve is used along with the stellar bulge and disk light profiles 
to decompose the mass distributions in NGC5005 and determine a dark matter halo model. The maximum stellar 
disk contribution to the total rotation curve is only about 70\%, suggesting that dark matter makes a
significant contribution to the dynamics at all radii.
\end{abstract}


\section{INTRODUCTION}
Understanding the role and significance of dark matter in the evolution of 
baryonic components (i.e., conversion of the gaseous disk into stars) is a critical aspect 
for realistic models of galaxy evolution. Observational studies aimed at addressing the 
influence of dark matter in galaxy formation and evolution often use mass decompostion to 
constrain the distribution of dark matter. Many of these projects utilize neutral hydrogen (HI)
synthesis observations 
to derive the kinematics for rotation curve decompostion (i.e., The HI Nearby Galaxy Survey (THINGS); 
\citealt{things}, Westerbork Survey of HI in Spiral Galaxies (WHISP); \citealt{whisp}). 
In this paper, we take advantage of the well-behaved outer disk
kinematics in the galaxy NGC5005 to perform a mass decomposition analysis. We use radio 
synthesis observations of HI and CO to probe both the gas surface density and dynamics as the 
gas follows the gravitational potential of the galaxy. The stellar contribution to the overall
mass budget of the galaxy is determined from deep near-infrared (NIR) observations. Finally, 
we use optical broadband B and R and narrowband H$\alpha$ observations to supplement the NIR and 
radio synthesis data to provide information about the dominant stellar populations and star 
formation activity.

Large projects like THINGS and WHISP have provided a foundation for numerous other studies 
investigating gas kinematics and morphology (e.g., \citealt{whispkinlopside},
\citealt{thingsmorph}, \citealt{thingssf}) and dark matter haloes (e.g., \citealt{thingsdmsim},
\citealt{dmthings}). However, an examination of the statistical significance of connections
between the dark matter and baryonic properties of these galaxies may be biased due to
the representative nature of the samples. 
In an effort to address fundamental questions regarding the growth and distribution of stellar 
disks in dark matter haloes in a statistical manner, we have undertaken a project 
correlating structural properties and star formation activity with the dark matter properties 
of the host galaxy. The project uses a statistical sample of 45 nearby galaxies and builds on
existing data obtained as part of the Extended Disk Galaxy Exploration Science (EDGES) Survey 
(\citealt{edges}), which includes deep 
{\it Spitzer Space Telescope} imaging observations. The statistical sub-sample consists of all galaxies 
in the EDGES Survey that are optimally suited for rotation curve decomposition, 
based primarily on the limited range of inclination angles that yield both accurate 
rotation curves and surface density profiles. Our combination of optical, 
NIR, and radio synthesis observations will enable a comprehensive statistical analysis of the 
evolution of galactic disks as a function of baryonic mass, environment, and dark matter mass.

In this paper, we focus our efforts on a single galaxy from the sample, NGC5005. This SAB(rs)bc
type galaxy has been studied extensively in the literature (e.g., \citealt{oldRCN5005}, 
\citealt{oldsurfphotN5005}, \citealt{oldradlightN5005}, \citealt{bulgeIFU}). It is known to 
host a low-ionization nuclear emission-line region (LINER) in its center (\citealt{linerref}) and 
has been classified as a low-luminosity active galactic nucleus (LLAGN) with a polycyclic aromatic 
hydrocarbon (PAH)-dominated mid-infrared spectrum by \cite{llagnref}. The morphology of NGC5005 as 
viewed in the infrared and optical features a strong bulge with boxy isophotes (\citealt{bulgemorph})
and a complicated dust morphology with no coherent structure (\citealt{dustmorph}). There is also
evidence of a fairly weak, round bar which is nearly aligned with the major axis of the galaxy.
Despite the weakness of the bar, \cite{bardyn} found that it could still be responsible for 
creating shocks visible in the central molecular gas velocity field. The molecular gas dynamics 
in NGC5005 were studied in more detail by \cite{COdynamics}. Their observations reveal CO 
emission from the nucleus, a ring at about 3 kpc, and a stream of gas northwest of the nucleus. The 
S-shape in the nuclear velocity field in \cite{COdynamics} is a signature of non-circular gas 
motions. \cite{COdynamics} cite a stellar bar of length $\sim$5 kpc as being accountable for 
infalling gas seen as the northwest stream and for creating the 3 kpc ring which may form at the 
ultraharmonic resonance of the bar or alternatively constitute a pair of spiral arms that originate 
at the bar's ends. A position-velocity cut along the major axis of NGC5005 reveals a steep rise in 
the rotational velocity of the CO emission (their Figure 5). The rotational velocity rises to 
approximately 300 km s$^{-1}$ within the central 10\arcsec.

NGC5005 lies in a relatively populated environment that is near the Ursa Major Cluster. 
The distance to NGC5005 is uncertain; distance measurements range
from 13.7 Mpc to 34.6 Mpc. In this paper, we adopt the Type Ia supernova distance of 16.5 Mpc 
(\citealt{distref}). NGC5005 is thought to be part of a physical galaxy pair with the nearby 
spiral galaxy NGC5033 (\citealt{N5033pair}). The two galaxies are separated by a projected distance 
of approximately 41\arcmin~($\sim$200 kpc at 16.5 Mpc) and are at similar redshifts. There are no 
clear signs of any interaction between the galaxy pair. There are two dwarf galaxies, 
SDSS J131115.77+365911.4 and SDSS J131051.05+365623.4, that are within 10\arcmin~of NGC5005 and 
also lie at similar redshifts. Farther out, there are many additional dwarf galaxies within one 
degree of NGC5005 which are part of the NGC5033 group. 

The goal of the present work is to investigate the dark matter and baryons in NGC5005. NGC5005 
is also a proof of concept for the larger study. The observational data are presented
in Section \ref{obsdata} and we focus on the kinematics in Section \ref{kin}. 
Section \ref{dmdecomp} describes the rotation curve decomposition analysis and the determination
of the dark matter distribution in NGC5005. Section \ref{conclusion} provides a brief summary of
the results.

\section{OBSERVATIONAL DATA \label{obsdata}}
\begin{figure}
\plotone{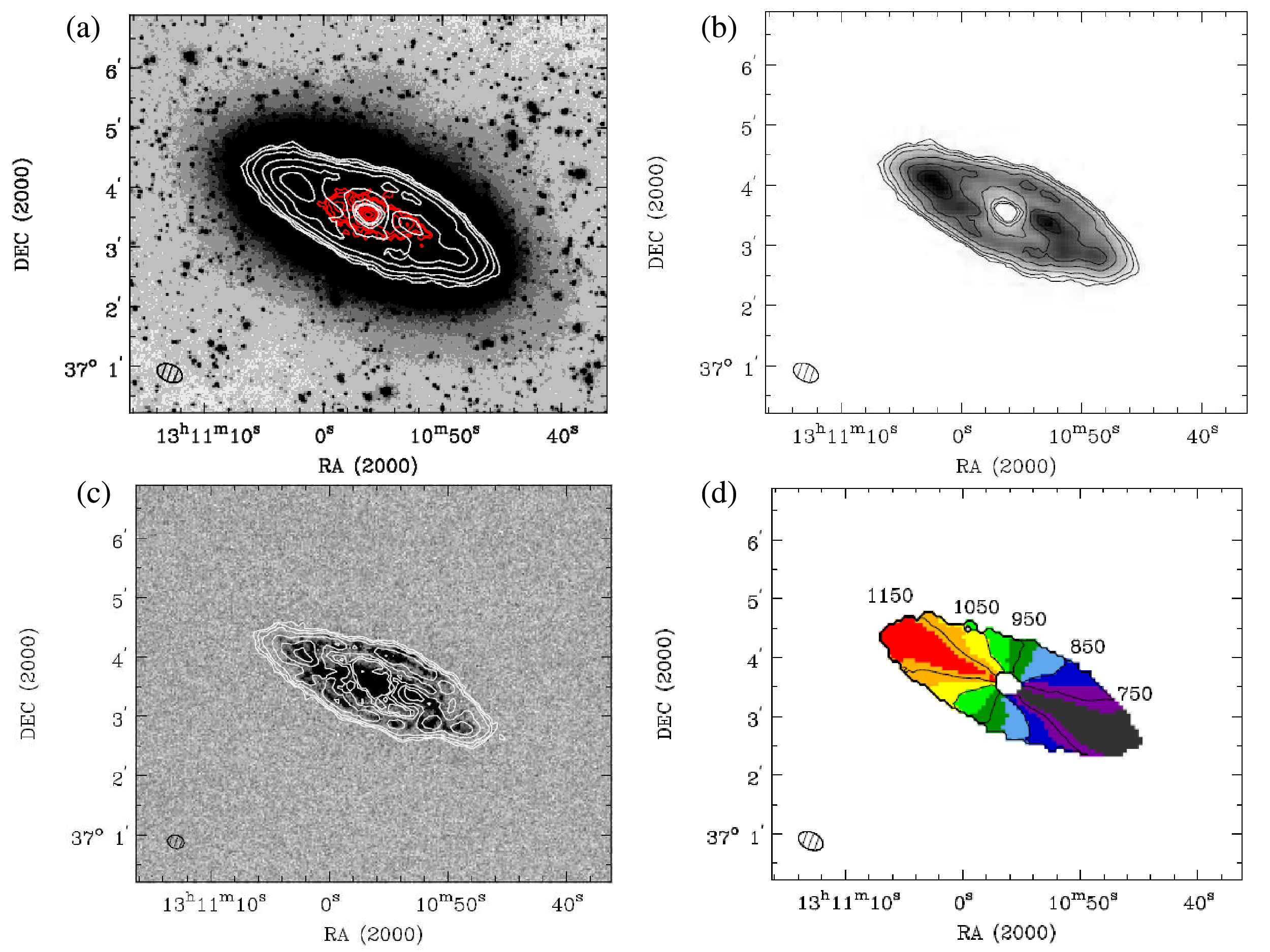}
\caption{(a) Low spatial resolution HI integrated intensity contours {\it (white)} and CO intensity 
contours from \cite{bimasongII} {\it (red)} overlaid on the {\it Spitzer} 3.6$\mu$m image. Note that 
this field of view focused in on NGC5005 does not include the entire low surface brightness stellar disk. 
(b) Low spatial resolution HI intensity contours overlaid on the low spatial resolution HI intensity 
image. Note the hole in HI emission in the center. (c) High spatial resolution HI intensity contours 
overlaid on the H$\alpha$ image. (d) HI velocity field, or first moment map with isovelocity contours
derived from the low spatial resolution HI data. The velocity contour values are 
indicated in the plot in units of km s$^{-1}$. The HI intensity contours represent column 
densities of 10$^{20}$ atoms cm$^{-2}$ $\times$ (1, 2, 4, 8, 12, 16) in images (a)-(c).\label{n5005}}
\end{figure}

Multiwavelength observations have been acquired to probe the stellar and gas content in NGC5005.
Deep NIR images taken at 3.6$\mu$m from the {\it Spitzer Space Telescope} trace the extended 
stellar population, while moderate depth optical broadband B and R and narrowband H$\alpha$ provide 
information about the dominant stellar population and star formation activity. Radio synthesis 
observations of neutral hydrogen from the Very Large Array (VLA)
\footnotemark\footnotetext{The Very Large Array is operated by the National Radio Astronomy 
Observatory, which is a facility of the National Science Foundation operated under 
cooperative agreement by Associated Universities, Inc.} are used to probe both the gas surface 
density and kinematics. Spectroscopic integral field unit (IFU) observations from SparsePak on 
the WIYN 3.5m telescope provide ionized gas kinematics in the central region of NGC5005. Figure 
\ref{n5005} shows the new VLA HI observations along with optical and NIR imaging data. Table 
\ref{intprops} provides a summary of integrated properties derived from this multifrequency dataset. 
We describe the data acquisition and processing below.

\begin{deluxetable}{lc}
\tabletypesize{\scriptsize}
\setlength{\tabcolsep}{25pt}
\tablecaption{Integrated Properties of NGC5005 \label{intprops}}
\tablewidth{0pt}
\tablehead{\colhead{Property} & \colhead{NGC5005}}
\startdata
Distance\tablenotemark{a} & 16.5 Mpc \\
Morphological Type\tablenotemark{b} & SAB(rs)bc \\
m$_{\rm B}$ & 10.74$\pm$0.02 \\
m$_{\rm R}$ & 9.29$\pm$0.02 \\
m$_{3.6}$ & 6.21$\pm$0.01 \\
M$_{\rm B}$ & -20.50$\pm$0.02 \\
(B-R)$_{0}$ & 1.43$\pm$0.02 \\
(B-3.6)$_{0}$ & 4.48$\pm$0.02 \\
(R-3.6)$_{0}$ & 3.05$\pm$0.02 \\
$\log$(H$\alpha$ flux) & -11.42$\pm$0.09 erg s$^{-1}$ cm$^{-2}$\\
Equivalent width & 7.93$\pm$0.32 \AA \\
Total SFR & 0.67$\pm$0.14 M$_{\sun}$ year$^{-1}$ \\
Total stellar mass\tablenotemark{c} & (9.16$\pm$1.83)$\times$10$^{10}$ M$_{\sun}$ \\
HI flux & 19.0$\pm$3.8 Jy km s$^{-1}$ \\
Total HI mass & (1.22$\pm$0.24)$\times$10$^9$ M$_{\sun}$ \\
CO flux\tablenotemark{b} & 1278$\pm$484 Jy km s$^{-1}$ \\
Total H$_2$ mass\tablenotemark{d} & (2.73$\pm$1.03)$\times$10$^9$ M$_{\sun}$ \\
Total dynamical mass\tablenotemark{e} & (2.82$\pm$0.47)$\times$10$^{11}$ M$_{\sun}$ \\
\enddata
\tablecomments{The apparent magnitudes are measured values and are not corrected for extinction.
The reported apparent magnitudes and colors are measured at R$_{25}$. M$_{\rm B}$ and colors are 
extinction corrected assuming A$_{\rm B}$=0.051 and A$_{\rm R}$=0.031 (\citealt{galextinct}). The extinction 
correction for the NIR is assumed to be negligible. The M$_{\rm B}$ value also includes a small correction 
(0.10 mag) based on an extrapolation of the observed stellar disk.}
\tablenotetext{a}{\cite{distref}}
\tablenotetext{b}{RC3; \cite{rc3}}
\tablenotetext{c}{Assumed best-fitting M/L$_{3.6}$ = 0.5$\pm$0.1}
\tablenotetext{d}{\cite{bimasongII}; mass corrected for distance=16.5 Mpc}
\tablenotetext{e}{Measured at R$_{\rm break}$, assuming the rotational velocity stays constant.}
\end{deluxetable}

\subsection{VLA Observations \label{vlaobs}}
Radio synthesis observations of NGC5005 using the 21 cm line of HI were obtained with the 
VLA in C configuration on 2013 August 4 and 
5 for a total on-source time of 407 minutes. The data had an initial frequency resolution of 
7.8125 kHz channel$^{-1}$ ($\sim$1.65 km s$^{-1}$ channel$^{-1}$). The standard flux calibrator 
3C286 was observed at the beginning and end of each observing block, and the phase calibrator 
J1227+3635 was observed approximately every 40 minutes, so that the data may be flux and phase 
calibrated.   

In order to flag uniformly, the data were loaded into AIPS \footnotemark\footnotetext{The
Astronomical Image Processing System (AIPS) has been developed by the NRAO.}
and the inner 75\%~of the bandpass was combined to generate a ``channel zero'' dataset. The
channel zero data was flagged and phase calibrated. The calibration solutions were applied
to the line data, which was then bandpass calibrated with 3C286. After calibrations, the data 
were continuum subtracted in the $uv$ plane and then 
combined. The combined, continuum subtracted datasets were imaged using the AIPS task IMAGR.
Three cubes were created with different Robust weighting parameters for varying
spatial resolution and a channel averaging of 3 for a resulting velocity resolution of 
$\sim$5 km s$^{-1}$ channel$^{-1}$ (for simplicity, these cubes are designated as low, medium, 
and high; Table \ref{cubeprops}). Only the cubes with the highest sensitivity (low) and highest 
resolution without severe loss of sensitivity (high) are shown in Figure \ref{n5005}. A lower 
velocity resolution cube (15 km s$^{-1}$ channel$^{-1}$) was additionally created to probe the 
lowest column density of HI gas. This velocity resolution was chosen to match the channel width 
to the expected line width from thermal gas motions. This cube was also smoothed spatially 
post-imaging to increase the sensitivity to low column density gas.

In addition to the new VLA C configuration data, there are existing archival VLA D configuration 
data taken on 2000 August 11 for a total on-source time of 80.3 minutes (project code AW536). These 
archival data were retrieved from the archive and processed in AIPS. Properties of the imaging of this 
data set are presented in Table \ref{cubeprops}. The archival D configuration data was 
not combined with the new C configuration data due to the lower velocity resolution and the smaller 
frequency coverage of the archival data. The galaxy signal nearly fills the bandpass in the D 
configuration observations. Despite this, the integrated HI flux and HI mass measured from the archival 
data is consistent with the new C configuration values within the measurement errors. Both this D configuration 
data and the spatially smoothed, 15 km s$^{-1}$ channel$^{-1}$  velocity resolution cube were used to 
confirm the lack of detection of low column density gas in the higher resolution cubes.

\begin{deluxetable}{lccccc}
\tabletypesize{\scriptsize}
\tablecaption{HI Synthesis Image Parameters \label{cubeprops}}
\tablewidth{0pt}
\tablehead{\colhead{Image Name} & \colhead{Velocity Resolution} & \colhead{Robust} & \colhead{Beam Size} 
& \colhead{Beam Position Angle} & \colhead{Noise} \\
\colhead{} & \colhead{(km s$^{-1}$)} & \colhead{} & \colhead{(arcsec)} & \colhead{(degrees)} & 
\colhead{(mJy beam$^{-1}$)}}
\startdata
low & 5.0 & 5 & 26.4$\times$17.2 & 64.2 & 0.57 \\
medium & 5.0 & 0.5 & 20.5$\times$14.8 & 67.0 & 0.61 \\
high & 5.0 & -0.5 & 16.9$\times$13.2 & 72.6 & 0.72 \\
smooth & 14.9 & 5 & 52.9$\times$34.5 & 64.2 & 0.53 \\
D config\tablenotemark{a} & 10.4 & 5 & 59.6$\times$51.8 & -45.2 & 0.96 \\
\enddata
\tablenotetext{a}{Archival VLA D configuration (project code AW536)}
\end{deluxetable}

The total intensity map and velocity field of the final combined C configuration data imaged at the 
low spatial and 5 km s$^{-1}$ channel$^{-1}$ velocity resolution are presented in 
Figures \ref{n5005} (a), (b), and (d). The velocity field
will be discussed in greater detail in Section \ref{hikin}. Figure \ref{n5005} (c)
shows the same data in contours imaged at high spatial resolution.
The first contour in Figures \ref{n5005} (a), (b), and (c) corresponds to a column density of 
10$^{20}$ atoms cm$^{-2}$. There are two notable features of the HI emission in NGC5005 visible in 
Figures \ref{n5005} (a) and (b). First, there 
is a slight hole or depression in the HI column density at the center. This is expected
to occur in galaxies where there is a phase change from atomic to molecular gas in the
central kiloparsec. Bar-driven transport also tends to concentrate the molecular gas in the
centers of galaxies. (\citealt{barCO}; see also \citealt{COdynamics}). 
Archival CO data allows us to trace the molecular gas component in the nucleus to fill in the
missing gas surface density information (Section \ref{codata}).
Second, the gaseous disk does not extend nearly as far as the stellar disk. Indeed, NGC5005 is 
known as an HI-deficient galaxy with a measured mass in HI that is almost half of its expected 
value (\citealt{HIdef}). Furthermore, we did not find low column density gas in either the lower velocity 
resolution, spatially smoothed map or in the archival D configuration data beyond what is shown in Figure 
\ref{n5005} (a).
The higher resolution HI contours displayed in Figure \ref{n5005} (c)
show that the neutral gas appears to be spatially coincident with the ionized gas seen in H$\alpha$,
as is expected for star forming regions in galaxies. Global properties of NGC5005 measured from 
this HI dataset are presented in Table \ref{intprops}. 

\begin{figure}
\plotone{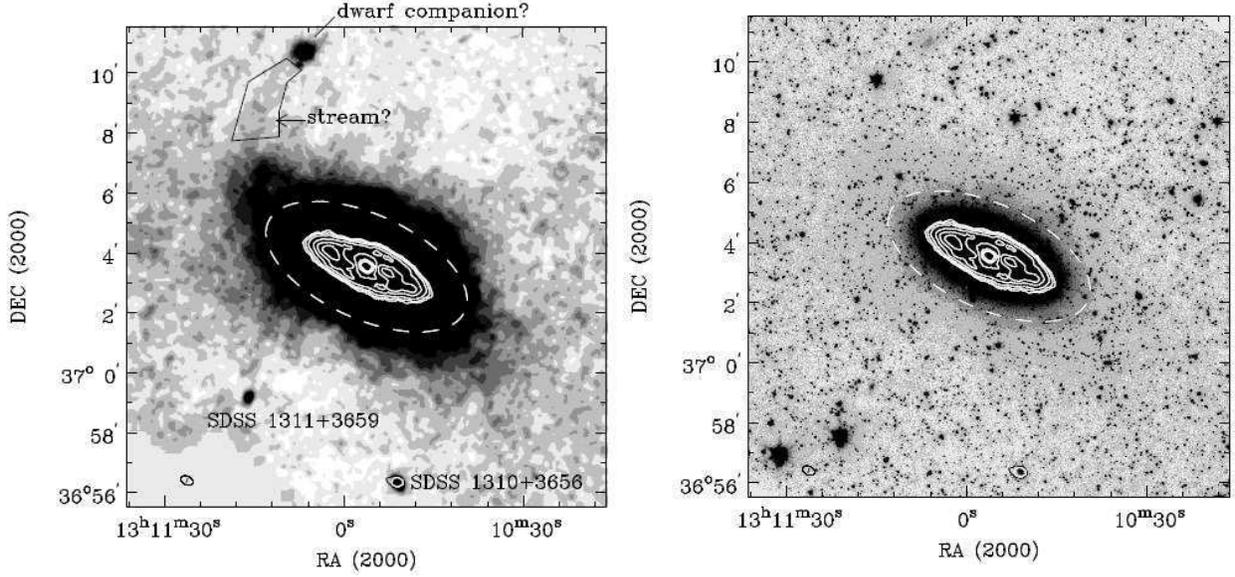}
\caption{Low spatial resolution HI integrated intensity contours {\it (white)} overlaid on the
{\it Spitzer} 3.6$\mu$m image to show the full extent of the stellar disk. The 3.6$\mu$m image 
on the left is smoothed and edited to highlight the low surface brightness features, 
including a possible 
dwarf companion to the north and a hint of a potential stellar stream connecting it to NGC5005. 
The outlined region demarcates where flux assumed to be part of the potential stream was measured. 
The image on the right shows the unmasked, unsmoothed large field of view around NGC5005. The dwarf 
companion SDSS J131051.05+365623.4 (labeled as SDSS 1310+3656 in the left panel) to the south 
of NGC5005 is detected in HI. The dwarf companion SDSS J131115.77+365911.4 (abbreviated
as SDSS 1311+3659) is also visible to the southeast of NGC5005 in the left panel. The white dashed ellipse 
indicates R$_{\rm break}$ (Section \ref{optdata}) in both images. \label{bigfov}}
\end{figure}

The field of view of the HI observations includes not only NGC5005, but also nearby dwarf galaxies 
SDSS J131051.05+365623.4, SDSS J131115.77+365911.4, SDSS J131105.57+371036.1, and 
SDSS J131126.81+371842.3. Only SDSS J131051.05+365623.4 is detected in the HI images (Figure 
\ref{bigfov}) with a measured flux of 0.21 Jy km s$^{-1}$ and line width at 50\%~of the peaks of
33 km s$^{-1}$; this yields a corresponding HI mass of 
1.33$\times$10$^{7}$ M$_{\sun}$ at the adopted distance of 16.5 Mpc. Outside of the primary beam, 
we also detect NGC5002 and NGC5033. No other known or suspected companions were detected in the 
HI observations.

\subsection{Archival CO \label{codata}}
The central hole in HI column density makes it difficult to extract trustworthy total
gas surface density measurements from the VLA data alone. Therefore, archival CO data from the 
BIMA Survey of Nearby Galaxies (BIMA SONG) (\citealt{bimasongII}) was acquired to supplement the 
HI observations of NGC5005.
Observations for the BIMA SONG survey were carried out from 1997 November through 
1999 December using the Berkeley-Illinois-Maryland Association (BIMA) millimeter interferometer in 
Hat Creek, CA. The archival CO data cube for NGC5005 has a velocity resolution of about 
10 km s$^{-1}$ and a spatial resolution of 6.2\arcsec~$\times$ 6.0\arcsec. The integrated CO intensity 
is mapped in red contours in Figure \ref{n5005} (a). A more detailed map can be found 
in \cite{bimasongII}. This map allows the central molecular gas surface density to be determined. 
Note that the spatial extent of the molecular gas is confined within the central $\sim$40\arcsec~or 
$\sim$3.2 kpc at 16.5 Mpc. As is typical for barred galaxies with CO emission, the molecular gas is 
concentrated in the center (\citealt{barCO}). In the present study, we focus on using the archival 
CO observations to improve the accuracy of the gas surface density measurements only. A complete 
analysis of the complex molecular gas dynamics in NGC5005 can be found in \cite{COdynamics}.

\subsection{WIYN SparsePak \label{spak}}
We obtained an H$\alpha$ velocity field for the central region of NGC5005 using the SparsePak
IFU (\citealt{spak}) on the WIYN 3.5m telescope. The SparsePak IFU is composed of eighty-two
5\arcsec~diameter fibers arranged in a fixed 70\arcsec~$\times$ 70\arcsec~rectangle. Observations were
acquired on the night of 26 April 2014 during reasonably clear, but non-photometric
weather with good seeing. We used the STA1 CCD binned 4 $\times$ 3 with the 316$@$63.4 
grating in eighth order, centered at 6686 \AA~to cover a wavelength range from 6480~\AA~to 6890~\AA. 
This resulted in a velocity resolution of 13.9 km s$^{-1}$ pixel$^{-1}$, which is sufficient to 
derive an accurate velocity field for NGC5005.

\begin{figure}
\plotone{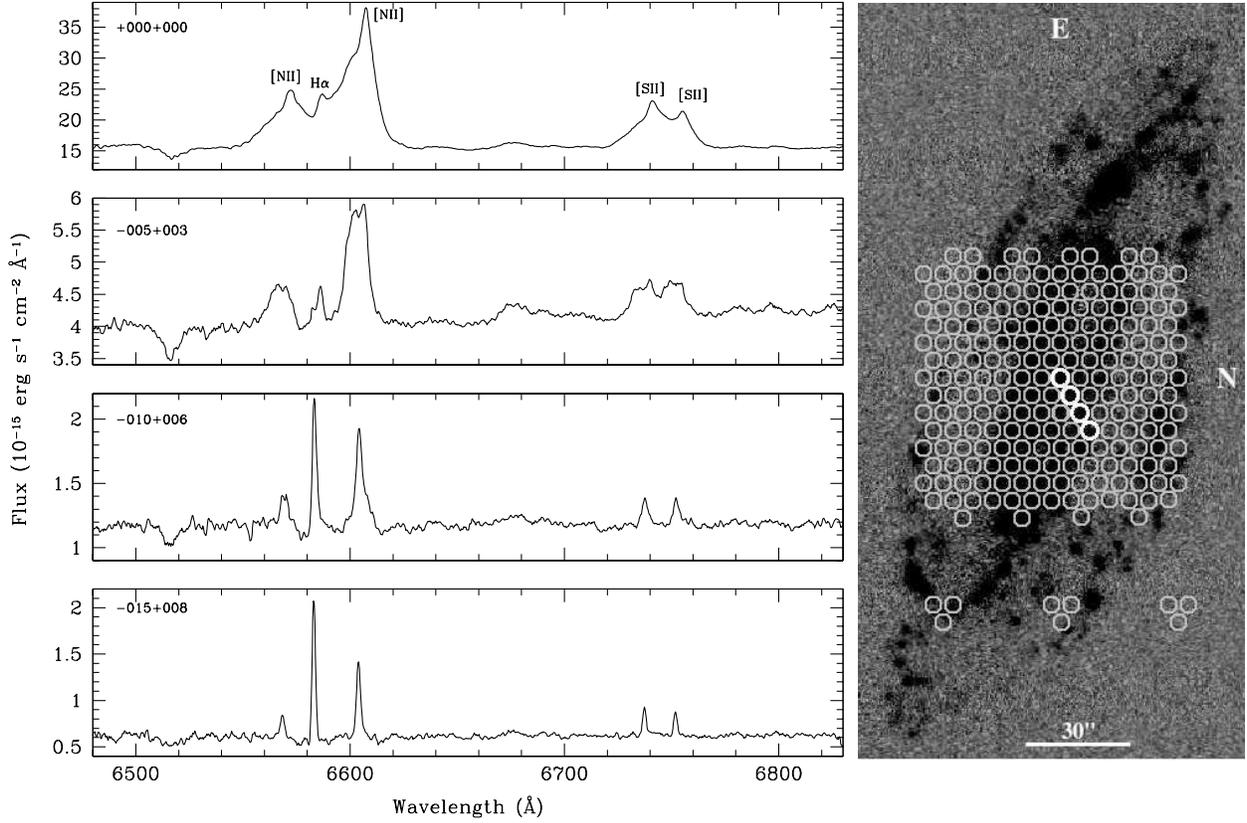}
\caption{Example spectra extracted from four different fibers from the SparsePak IFU.
The positions of the SparsePak fiber apertures on NGC5005 are displayed on the 
narrowband H$\alpha$ image from HDI on the right. The fibers from which the four spectra
were extracted are highlighted in white. The image has been rotated such that
north is to the right and east is up. The numbers in the top left corner of each spectrum 
panel indicate the fiber's position offset from the central fiber in arcseconds. \label{spectra}}
\end{figure}

For simplicity, the SparsePak array was aligned on the sky with a position angle of 0\degr. 
We used a three pointing dither pattern to spatially fill in gaps between fibers. We took
three exposures at 900 s for each dither pointing in order to be able to detect diffuse 
ionized gas, not just star forming regions. Additional observations of blank sky were also
taken to remove sky line contamination more accurately, as NGC5005 is much more extended 
than the SparsePak field-of-view (see Figure \ref{spectra}). 

The SparsePak data was processed using standard tasks in the HYDRA package within IRAF.\footnotemark 
\footnotetext{IRAF is distributed by NOAO, which is operated by the Association of Universities for 
Research in Astronomy, Inc., under cooperative agreement with the National Science Foundation.} The data
were bias-subtracted and flattened, and the IRAF task DOHYDRA was used to fit and extract apertures
from the IFU data. The spectra were wavelength calibrated using a wavelength solution created
from ThAr lamp observations. The three exposures for each dither pointing were cleaned
of cosmic rays. The individual images were sky subtracted using the separate sky pointing scaled to the 
strength of the 6577 \AA~ sky line, which is close to the redshifted H$\alpha$ and [NII] $\lambda$6584 
lines in some locations. 
The cleaned, sky subtracted images were averaged together to increase the signal-to-noise ratio and then 
flux calibrated using observations of spectrophotometric standards from \cite{spakstandards}.

We illustrate representative spectra from four SparsePak fibers in the central 
region of NGC5005 in Figure \ref{spectra} to demonstrate the types of spectral emission found
in NGC5005. The locations of these fibers start in the
central nucleus of NGC5005 and move out towards the northwest just off the minor axis (see the
rotated H$\alpha$ image on the right). While our primary analysis relies on 
cross-correlation of the entire spectrum, we also fit Gaussian profiles to emission lines of H$\alpha$, 
[NII], and [SII] in all the spectra to estimate the full width half maximum (FWHM) values and the emission
line ratios. As seen previously (\citealt{broadline}) the strength of the [NII] $\lambda$6584 
line emission relative to H$\alpha$ in the central fiber (+000+000) implies
non-thermal emission processes (\citealt{bpt}). The broadness of the nuclear emission lines 
from the center of NGC5005 are suggestive that the nucleus is accretion powered (\citealt{linerref}). 
We find an approximate FWHM of 900$\pm$200 km s$^{-1}$ based on single Gaussian fits
to [NII] and [SII] lines within the central 5\arcsec~fiber.
The second spectrum in Figure \ref{spectra} (-005+003) was extracted from the adjacent fiber to the 
northwest of the central fiber. This spectrum displays double-peaked emission lines that suggests either 
infalling or outflowing gas. 
Indeed, this fiber is placed near where \cite{COdynamics} detected infalling molecular gas in their 
CO observations of NGC5005. The third fiber's spectrum (-010+006) still shows broad emission lines
with non-thermal line strengths, but less so in both aspects than the central fiber. 
This spectra is representative of diffuse ionized gas emission in NGC5005,
which we measure to have typical FWHMs of about 140$\pm$10 km s$^{-1}$ and [NII] $\lambda$6584/H$\alpha$ 
line ratios of 3.0$\pm$1.2, which is consistent with non-thermal or shock excited gas.
The fourth fiber's
spectrum (-015+008) shown in the bottom panel of Figure \ref{spectra} is placed on the ring of 
H$\alpha$ emission surrounding the nucleus. This spectrum displays narrower line widths and stronger 
H$\alpha$ to [NII] $\lambda$6584 line emission strengths. We find typical FWHMs of approximately
80$\pm$5 km s$^{-1}$ and [NII] $\lambda$6584/H$\alpha$ ratios of 0.4$\pm$0.1 in similar spectra, which are 
indicative of primarily thermal emission processes most likely due to recent star formation activity 
(\citealt{bpt}).

Inspection of spectra extracted from a few other fibers revealed
additional locations with double-peaked emission lines, similar to the -005+003 fiber spectrum.
These spectra were located approximately east-west on the sky along where evidence from \cite{COdynamics} 
suggests a stellar bar exists. We note, however, that double-peaked emission lines may also
occur due to smearing of the rotation curve within the 5\arcsec~SparsePak fibers. The spectra
along the major axis would be most susceptible to this beam smearing as the velocity gradient
is highest here. Since the bar is oriented close to the major axis of NGC5005, it would be 
difficult to verify that the double-peaked emission lines are signatures of infalling gas along
the bar. Alternatively, the beam smearing effect would be minimized for velocities along
the minor axis, so it is reasonable to assume that the double-peaked emission lines seen in the -005+003
spectrum in Figure \ref{spectra} are real and are not merely due to observational effects.

The flux calibrated, sky subtracted spectra were cross-correlated with a template emission
line spectrum to extract the luminosity weighted mean recessional velocity at every 
position using the IRAF task FXCOR. 
We used the spectrum of an HII region from within NGC5005 as our reference template for the 
cross-correlation analysis due to its strong narrow emission lines. Spectra with a
Tonry and Davis ratio (TDR; \citealt{tdr}) less than 5 were not included in deriving the velocity field.
Spectra with broad or double-peaked emission line profiles resulted in broad cross-correlation peaks
that were nonetheless of high significance and had a well-defined centroid.
The velocity values from FXCOR were then placed into a grid that mapped the SparsePak fiber locations. 
To fill in the missing spacings, the velocity field was then interpolated using the
average value from the nearest eight pixels. Analysis of the velocity field will be discussed
further in Section \ref{spakkin}.

\subsection{Spitzer 3.6$\mu$m Data \label{3.6data}}
Imaging observations at 3.6$\mu$m were obtained using the Infrared Array Camera (IRAC) on the
{\it Spitzer Space Telescope} on 2012 July 12 and 19. Observations were conducted in a grid-like 
mapping pattern of 10 $\times$ 9 dithered pointings, where each pointing was offset by 
$\sim$100\arcsec~to cover a total area of 18\arcmin~$\times$ 33\arcmin. Observations were taken in 
two sets separated by several days to enable asteroid removal from the mosaicked images. The 
resulting map has a total exposure time of 1800 s pixel$^{-1}$. This exposure time provides a 
sensitivity of a few$\times$0.01 M$_{\sun}$ pc$^{-2}$ over a field of view 5 times the size of 
the optical radius R$_{25}$. After standard processing, the individual pointings were combined
into maps using the MOsaicker and Point source EXtractor (MOPEX) software. Images were drizzled to resample 
the data to 0.75\arcsec~pixel$^{-1}$ (\citealt{drizzle}). The maps were fit for a first 
order sky subtraction to remove any gradient in the sky level. 

Figure \ref{bigfov} shows the full extent of the faint stellar disk component of NGC5005 detected 
in the final mosaicked and sky subtracted 3.6$\mu$m image. The image on the left has been smoothed 
to bring out the lowest surface brightness features. The immense sensitivity over such a large field 
of view enables a complete study of the faint extended stellar component of NGC5005. NGC5005 has 
a large high surface brightness disk, which is surrounded by an even more extended low surface 
brightness component. This faint stellar population extends about twice as far as the break or transition
radius between the high and low surface brightness components and nearly three times farther than the gas disk 
detected in HI. This is an unusual result, as most spiral and irregular galaxies with HI have a
diameter in HI that is larger than the optical diameter (\citealt{DHID25}). 

There is also a feature that appears to be a faint stellar stream extending to the 
northeast towards an apparent companion galaxy SDSS J131105.57+371036.1. 
To measure the stellar mass associated with the apparent stellar stream, we  
used the smoothed and masked image to create a hand-drawn polygonal aperture 
that traces the area associated with the diffuse light (Figure \ref{bigfov}). We measured a lower limit
for the flux in the stream of 0.15 $\pm$ 0.13 mJy based on a conservatively 
masked image. However, this value significantly underestimates the total 
flux associated with the stream due to the significant masking required to
remove the large area associated with the bright foreground star located 
in the middle of the stream. Thus, to obtain a more accurate estimate of 
the total mass associated with the stream, we replaced the masked pixels with 
a value interpolated from neighboring pixels and remeasured the total flux to 
be 0.35 mJy. Assuming the best-fitting disk mass-to-light ratio 
of 0.5 (Section \ref{dmdecomp}) and a distance to NGC5005 of 16.5 Mpc, this corresponds 
to a total stellar mass for the stream of 1.8 $\times$ 10$^7$ M$_{\odot}$, which is an 
order of magnitude less massive than notable streams identified around other galaxies, such as M63 
(\citealt{M63stream}, \citealt{ShawnM63}) and M83 (\citealt{KateM83}).
The apparent companion galaxy is not detected in the HI observations and does
not have a redshift measurement available in the literature. However,
if SDSS J131105.57+371036.1 is at the same distance as NGC5005 and we again assume a mass-to-light
ratio of 0.5, it would have a total stellar mass of 2.3~$\times$~10$^7$~M$_{\odot}$ (Table
\ref{otherprops}). Given the slightly larger mass of the apparent companion galaxy,
interactions between the apparent dwarf companion and NGC5005 cannot be immediately
ruled out as an origin for the potential stellar stream.

\begin{deluxetable}{lcccccrccccc}
\tabletypesize{\scriptsize}
\setlength{\tabcolsep}{3pt}
\rotate
\tablecaption{Properties of Known and Suspected Companions to NGC5005 \label{otherprops}}
\tablewidth{0pt}
\tablehead{\colhead{Name} & \colhead{RA, Dec} & \colhead{Systemic\tablenotemark{a}} &
\colhead{Angular} & \colhead{Physical\tablenotemark{b}} & \colhead{Size\tablenotemark{c}} & 
\colhead{Position} & \colhead{m$_{\rm B}$} & 
\colhead{m$_{\rm R}$} & \colhead{m$_{3.6}$} & \colhead{(B-R)$_0$} & \colhead{M$_{\rm B}$\tablenotemark{b}} \\
\colhead{} & \colhead{(J2000)} & \colhead{Velocity} & 
\colhead{Separation} & \colhead{Separation} & \colhead{(a$\times$b)} & \colhead{Angle} & \colhead{} & 
\colhead{} & \colhead{} & \colhead{} & \colhead{} \\
\colhead{} & \colhead{} & \colhead{(km s$^{-1}$)} & \colhead{(arcmin)} & \colhead{(kpc)} & 
\colhead{(\arcsec$\times$\arcsec)} & \colhead{($\degr$)} & \colhead{} & \colhead{} & \colhead{} & 
\colhead{} & \colhead{}}
\startdata
NGC 5005 & 13:10:56.2, 37:03:33 & 948 & - & - & 336.4$\times$162.4 & 65.55 &
10.74$\pm$0.02 & 9.29$\pm$0.02 & 6.21$\pm$0.01 & 1.43$\pm$0.02 & -20.50$\pm$0.02 \\
SDSS J131115.77+365911.4 & 13:11:15.7, 36:59:14 & 956 & 5.85 & 28.1 & 42.4$\times$21.6 & -22.24 & 
18.56$\pm$0.07 & 17.53$\pm$0.07 & 15.75$\pm$0.01 & 1.01$\pm$0.09 & -12.58$\pm$0.07 \\
SDSS J131051.05+365623.4 & 13:10:51.0, 36:56:23 & 1026 & 7.24 & 34.7 & 43.8$\times$28.2 & 32.21 &
17.76$\pm$0.05 & 16.80$\pm$0.05 & 14.81$\pm$0.01 & 0.94$\pm$0.07 & -13.38$\pm$0.07 \\
SDSS J131105.57+371036.1 & 13:11:06.2, 37:10:40 & - & 7.29 & 35.0 & 62.8$\times$45.4 & -62.08 &
17.71$\pm$0.10 & 16.75$\pm$0.10 & 15.15$\pm$0.01 & 0.94$\pm$0.14 & -13.43$\pm$0.10 \\
SDSS J131058.75+364943.8 & 13:10:58.6, 36:49:42 & - & 13.72 & 65.9 & 59.4$\times$36.2 & -49.14 & 
18.28$\pm$0.12 & 17.29$\pm$0.12 & - & 0.97$\pm$0.17 & -12.86$\pm$0.12 \\
SDSS J131058.04+364812.4 & 13:10:58.1, 36:48:12 & - & 15.35 & 73.7 & 39.6$\times$25.0 & -8.15 & 
19.46$\pm$0.13 & 18.25$\pm$0.13 & - & 1.19$\pm$0.19 & -11.67$\pm$0.13 \\
SDSS J131126.81+371842.3 & 13:11:26.9, 37:18:42 & 960 & 16.33 & 78.4 & 63.0$\times$50.8 & -7.90 &
17.21$\pm$0.06 & 16.11$\pm$0.06 & - & 1.08$\pm$0.09 & -13.93$\pm$0.06 \\
kkh 081 & 13:11:11.6, 36:40:46 & 1032 & 22.94 & 110.1 & 58.6$\times$35.8 & 17.85 &
18.70$\pm$0.16 & 17.68$\pm$0.16 & - & 1.00$\pm$0.24 & -12.44$\pm$0.16 \\
NGC 5002 & 13:10:37.9, 36:38:03 & 1091 & 25.78 & 123.7 & 212.8$\times$114.4 & -6.64 &
14.16$\pm$0.04 & 13.33$\pm$0.04 & - & 0.81$\pm$0.05 & -16.98$\pm$0.04 \\
\enddata
\tablecomments{The apparent magnitudes are measured values and are not corrected for extinction.
B-R and M$_{\rm B}$ are extinction corrected assuming A$_{\rm B}$=0.051 and A$_{\rm R}$=0.031 
(\citealt{galextinct}). The extinction correction for the NIR is assumed to be negligible.}
\tablenotetext{a}{From NED, if known (http://ned.ipac.caltech.edu/)}
\tablenotetext{b}{Assuming a distance of 16.5 Mpc}
\tablenotetext{c}{Sizes indicate the diameters in arcseconds of the apertures used to measure the reported magnitudes,
except for NGC5005 which is the size measured at the optical diameter D$_{25}$.}
\end{deluxetable}

\begin{figure}
\plotone{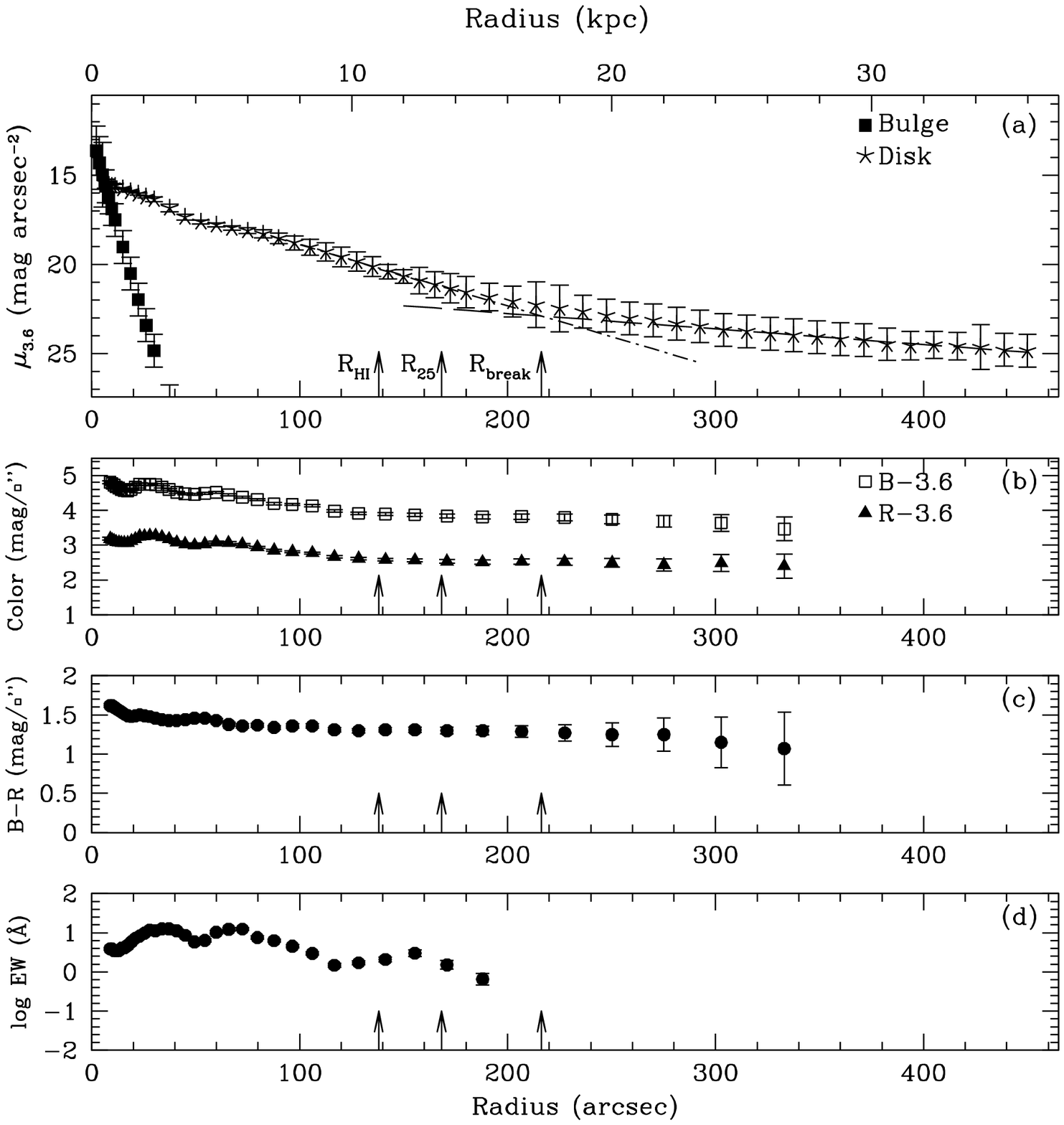}
\caption{(a) Surface brightness profile at 3.6$\mu$m decomposed into bulge ({\it squares}) and 
disk ({\it stars}) components. The dashed-dotted and dashed lines are exponential fits 
to the radial ranges 80\arcsec-180\arcsec~and 280\arcsec-443\arcsec, respectively. The arrows 
represent the radial extent of the HI disk, the
optically defined R$_{25}$, and the approximate radius at which there is an observed change in
slope of the outer exponential disk. (b) Radial trend in B-3.6 ({\it squares}) and R-3.6 ({\it triangles}) 
color. (c) Radial trend in B-R color. (d) Radial trend in the $\log$ H$\alpha$ equivalent width. 
\label{phot}}
\end{figure}

The 3.6$\mu$m image of NGC5005 was decomposed into bulge and disk components to represent the 
stellar mass distributions in the rotation curve decomposition analysis. Note that the emission
at 3.6$\mu$m is predominantly stellar, and not greatly affected by PAH emission at 3.3$\mu$m 
(\citealt{3.6noPAH}). We used {\sc DiskFit}
(\citealt{diskfit}) to derive a photometric model that fit bulge and disk components to the
3.6$\mu$m image itself. We did not include a bar component in the photometric model, despite evidence
for its existence (\citealt{bardyn}, \citealt{COdynamics}), since we only fit stellar bulge
and disk components in the mass decomposition (Section \ref{dmdecomp}). Our models are not
sensitive to the exact inner structure of the galaxy, so the lack of a bar component will have
a negligible effect on the derived dark matter profile. Radial surface brightness 
profiles of the resulting bulge and disk models are shown in Figure \ref{phot} (a). We additionally 
extracted a surface brightness profile from the 3.6$\mu$m image using concentric, con-eccentric 
ellipses to provide a consistency check. The {\sc DiskFit} model and the ellipse photometry yielded 
similar total surface brightness profiles. NGC5005's surface brightness profile displays a relatively 
small bulge component with a complex disk component featuring a break, or change in slope of the
exponential disk where the transition from high to low surface brightness stellar disk occurs. 
The break radius was estimated by fitting two exponentials to the radial ranges of
80\arcsec-180\arcsec~(inner) and 280\arcsec-443\arcsec~(outer). The radial
ranges were chosen so as to avoid locations where the surface brightness profile displays a
change in slope. Specifically, the inner limit of 80\arcsec~avoids the bulge component and the dip
in surface brightness around 40\arcsec. Scale lengths of these inner and
outer exponential fits are given in Table \ref{radprops}. The break radius is defined as the
radius at which the two different exponential fits intersect, which is 216.3\arcsec~(or
17.3 kpc at 16.5 Mpc). This break radius is indicated on the bottom axis of the panels in
Figure \ref{phot}, along with the optically defined R$_{25}$ (see Section \ref{optdata}) and radial 
extent of the HI disk. The size of the neutral gas disk is only about 60\%~of the size of the high
surface brightness stellar disk, as measured by R$_{\rm break}$. Figure \ref{phot} (b) shows the radial 
trends in B-3.6 and R-3.6 colors. The colors display nearly identical trends with radius featuring
a relatively complex distribution in the inner $\sim$80\arcsec~most likely due to star formation
in spiral arms.

\begin{deluxetable}{lc}
\tabletypesize{\footnotesize}
\tablecaption{Radial Properties of NGC5005 \label{radprops}}
\tablewidth{0pt}
\tablehead{\colhead{Parameter} & \colhead{Value}}
\startdata
Inner scale length (3.6$\mu$m) & 31.6\arcsec~(2.5 kpc at 16.5 Mpc) \\
Outer scale length (3.6$\mu$m) & 124.6\arcsec~(10.0 kpc at 16.5 Mpc) \\
Outer break radius (3.6$\mu$m) & 216.3\arcsec~(17.3 kpc at 16.5 Mpc) \\
D$_{25}$ & 336.4\arcsec~(26.9 kpc at 16.5 Mpc) \\
HI size at 10$^{20}$ atoms cm$^{-2}$ & 276\arcsec~(22.1 kpc at 16.5 Mpc) \\
Concentration (C=5$\log{\frac{R_{80}}{R_{20}}}$) & 4.3 \\
$\log$(Equivalent width) gradient & -0.11 $\log{(\rm EW)}$ arcmin$^{-1}$ \\
Color gradient (B-R) & -0.13 color arcmin$^{-1}$ \\
\enddata
\end{deluxetable}

Aperture photometry on the 3.6$\mu$m image was carried out using the GALPHOT\footnotemark 
\footnotetext{GALPHOT is a collection of scripts in the IRAF STSDAS environment first developed by 
W. Freudling and J. J. Salzer. The current version has been further enhanced by members of the Cornell 
Extragalactic Group and is maintained by M. P. Haynes.} package for IRAF. Concentric, con-eccentric
ellipses were used to measure the integrated magnitude at R$_{25}$ (Section \ref{optdata}) given in
Table \ref{intprops}. No foreground extinction correction was used for the NIR since the expected
value is negligible. The total stellar mass 
presented in Table \ref{intprops} was determined from the total stellar luminosity at 3.6$\mu$m assuming 
a mass-to-light ratio (M/L) of 0.5 (see Section \ref{dmdecomp}). The luminosity was calculated from the 
absolute magnitude at 3.6$\mu$m, which includes a slight correction (0.04 mag) based on an extrapolation 
of the 3.6$\mu$m surface brightness. The ellipse photometry was also used to estimate the concentration
of light in NGC5005 using the definition from \cite{Ccomp} (see Table \ref{radprops}). Our calculated
value of 4.3 is towards the higher end of concentrations and is in agreement with the value of 4.16
measured in the R-band for NGC5005 by \cite{Ccomp}. \cite{CVirgo} performed a study of NIR surface brightness 
distributions of Virgo cluster galaxies and discovered a bimodality of concentration with peaks at 
C = 3 and 4.5 in high surface brightness disk galaxies. The concentration value determined here for
NGC5005 is consistent with the second peak in the bimodal distribution. 

\subsection{Optical Data \label{optdata}}
Optical broadband B and R and narrowband H$\alpha$ images provide insight into changes in the 
age or metallicity of the dominant stellar population in NGC5005. Observations were taken with 
the WIYN 0.9m telescope at Kitt Peak National Observatory in February 2013 with the S2KB imager
and April 2014 with the Half Degree Imager (HDI). NGC5002, which was detected outside the primary
beam in the VLA data, was also observed with HDI. During both observing runs, narrowband imaging 
was done using a filter with FWHM of about 60 \AA~centered at a wavelength 
of 6580 \AA. A filter with similar FWHM and slightly offset central wavelength was used for continuum 
subtraction. The H$\alpha$ images were taken with total exposure times of 2 $\times$ 20 minutes in the 
6580 \AA~filter, and 20 minutes in the narrowband continuum filter. Broadband imaging was done with three 
exposures of 900 and 300 s with the B and R filters, respectively. An additional short R-band exposure 
was acquired to replace saturated pixels in the core of NGC5005. 

The optical images were reduced and analyzed with the IRAF package. The image
reduction included bias subtraction and flat fielding. Both the S2KB and HDI images were flat-fielded
using twilight flats for the broadband filters and dome flats for the narrowband. Prior to combination,
sky values were measured as the mode of galaxy-free regions of the images. The images were then scaled, 
aligned, and averaged together to make a final combined image. The WCS was updated using star lists 
from the USNO catalog. The S2KB and HDI images were processed independently
in order to provide internal consistency checks for the derived structural parameters. Since the HDI
images have slightly higher quality and a larger field of view than the S2KB images, the subsequent
analysis will focus on results from the HDI observations.

In addition to NGC5005, our HDI observations include eight known or suspected companion galaxies,
including NGC5002 (Table \ref{otherprops}). Velocities presented for the known companions in Table
\ref{otherprops} are from the NASA/IPAC Extragalactic Database (NED).\footnotemark \footnotetext{The 
NASA/IPAC Extragalactic Database (NED) is operated by the Jet Propulsion Laboratory, California Institute 
of Technology, under contract with the National Aeronautics and Space Administration.} The three 
unconfirmed companions are diffuse, irregular galaxies that do not appear to be larger background galaxies
based on visual inspection of the broadband images. These galaxies are identified using the 
Sloan Digital Sky Survey (SDSS) designation that most closely matched the photometric center of the galaxies.
No corresponding HI or H$\alpha$ emission was detected for the unconfirmed companion galaxies that look 
like they might be associated with NGC5005. The absolute B-band magnitudes presented in Table 
\ref{otherprops} are calculated assuming a distance of 16.5 Mpc.

Aperture photometry of NGC5005 using concentric, con-eccentric ellipses was carried out to 
derive surface brightness magnitudes for both the B- and R-band HDI images.
Table \ref{intprops} gives measured integrated magnitudes derived from the aperture photometry. The
integrated apparent magnitudes tabulated here are measured at the radius at 
which the B-band surface brightness equals 25 mag arcsec$^{-2}$ (R$_{25}$) and are not corrected for
foreground extinction. The absolute B-band magnitude has been corrected for extinction (\citealt{galextinct})
and includes a correction (0.10 mag) based on an extrapolation of the B-band surface brightness profile. 
Integrated colors are also measured at R$_{25}$ and have been extinction corrected.  

Inspection of the H$\alpha$ image indicates that there are no visible star formation 
regions beyond R$_{25}$. Aperture photometry was performed on the H$\alpha$ image to measure an 
H$\alpha$ flux and derive an equivalent width.  
The current total star formation rate (SFR) is derived from the H$\alpha$ flux (measured at
R$_{25}$) using the calibration given in \cite{sfrcal}. The calculated SFR of 0.67$\pm$0.14 
M$_{\sun}$ yr$^{-1}$ is
much lower than the median value of about 1.5 M$_{\sun}$ yr$^{-1}$ found by \cite{HaGS} for galaxies of
the same Hubble T-type as NGC5005. However, this SFR is most likely an underestimate due to internal
extinction, which we cannot accurately account for without a spatially resolved total infrared map due
to the strong active galactic nucleus (AGN) contamination.
The equivalent width (EW) of the H$\alpha$ emission line is used as a tracer of the specific star formation
rate. It is calculated by dividing the H$\alpha$ flux by the continuum flux density measured from the 
H$\alpha$ narrowband continuum filter. It, therefore, serves as an indicator of the strength of the 
current SFR relative
to the past average SFR. A larger EW value would indicate a larger current SFR relative to the continuum,
or past average star formation. NGC5005 has a relatively small EW value of 7.93$\pm$0.32 \AA, which is
consistent with the little H$\alpha$ emission seen in the narrowband image. This EW is also more of an
upper limit as it is contaminated by strong [NII] emission from the AGN in NGC5005. Despite this, the
derived EW value places NGC5005 in the same regime as other Sbc morphological type galaxies
in the M$_{\rm B}$-EW and (V$_{\rm max}$)-EW planes as found in \cite{11HUGS}.

In addition to integrated properties from the aperture photometry, radial profiles were also created to 
examine the surface brightness values in each elliptical annulus. Figure \ref{phot} shows the radial changes 
in the surface colors B-3.6, R-3.6 ({\it b}), and B-R ({\it c}) (see also Section \ref{3.6data}), as well as
the radial trend in EW ({\it d}). The radial trend in B-R is typical of large spiral galaxies, going from
redder in the bulge-dominated nuclear region to bluer in the disk where more of the current star formation
is occurring (e.g., \citealt{galcolors}). After the initial gradient from red to blue, the B-R color remains 
roughly constant through the disk of NGC5005. 
There does not appear to be any change in B-R color corresponding to either the extent
of the HI gas, R$_{25}$, or the change in the outer disk exponential slope (R$_{\rm break}$). The radial trend 
in EW is much less regular than the B-R color trend. The higher values of EW correspond to spiral arms 
where there are more HII regions from recent star formation activity. 
Table \ref{radprops} gives estimates of the EW gradient and B-R color gradient from a linear fit to
these profiles. 

\begin{figure}
\plotone{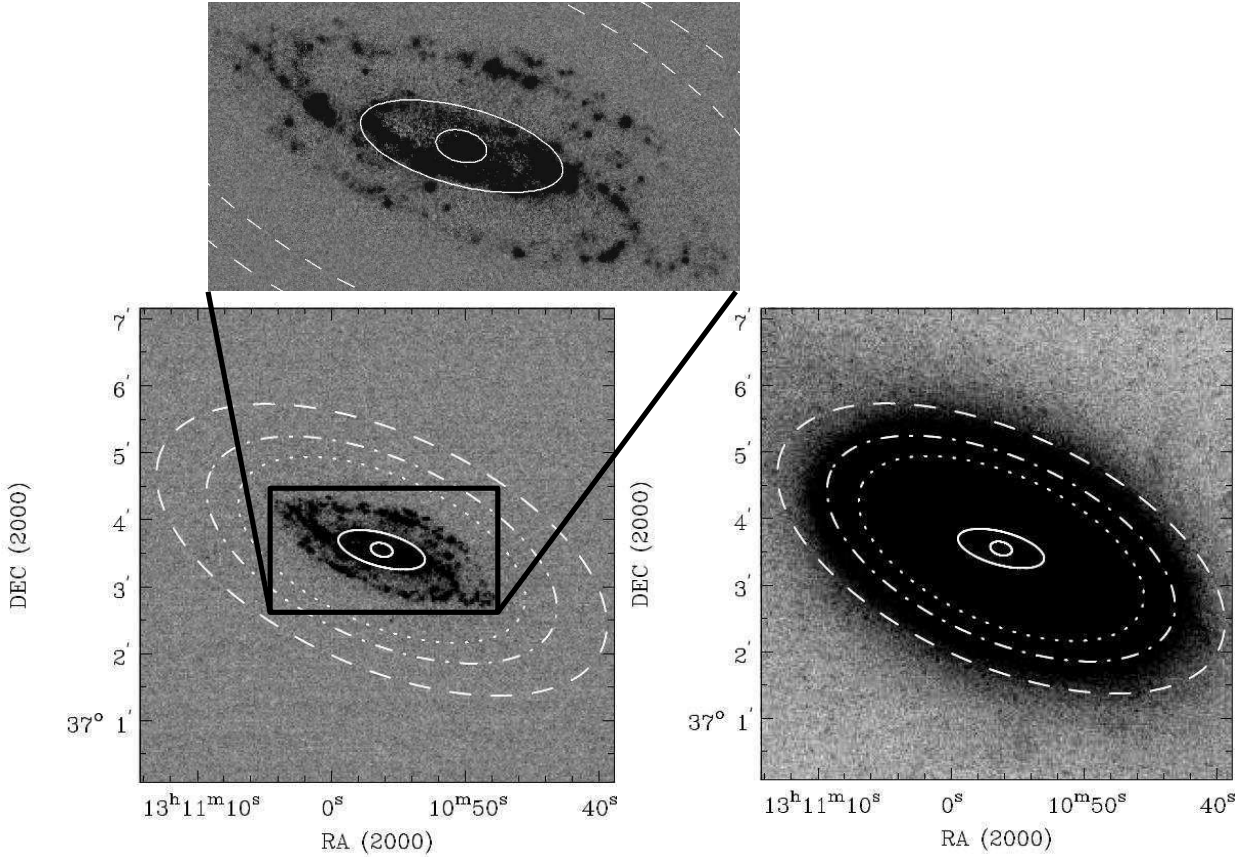}
\caption{{\it Left}: Narrowband H$\alpha$ image of NGC5005 from HDI. {\it Right}: Masked and interpolated
R-band image of NGC5005 from HDI. The inner two solid ellipses demonstrate the size of the nuclear disk and 
ring structure, respectively (see Section \ref{optdata}). The outer three ellipses demarcate the three radii 
shown in Figure \ref{phot}: R$_{\rm HI}$ ({\it dotted}), R$_{25}$ ({\it dashed-dotted}), and R$_{\rm break}$ 
({\it dashed}), listed in order of increasing size. \label{ellipses}}
\end{figure}

For visual reference, the characteristic radii described above are overplotted on both
the H$\alpha$ and R-band images in Figure \ref{ellipses}. The images are displayed 
with the same scale to exemplify the concentration of the ionized gas emission ({\it left}) and extent of 
the stellar disk ({\it right}). As is expected, the morphology of the ionized gas emission in the central 
$\sim$40\arcsec~directly matches the morphology of molecular gas emission mapped in CO by \cite{COdynamics}. 
As can be seen in the zoom-in image, this morphology features a central nuclear disk 
(first solid ellipse) with a ring of emission at about 30\arcsec-40\arcsec~(second solid ellipse; referred 
to as the 3 kpc ring in \citealt{COdynamics}). The two central ellipses have semi-major axes of 
10\arcsec~and 40\arcsec, respectively, assuming an inclination angle of 53\degr~(from \citealt{COdynamics}) 
and position angles of 75\degr. The H$\alpha$ emission continues beyond this central region, tracing the 
spiral arms which appear to emanate from either end of the ring along the semi-major axis. The ionized gas 
emission is contained within the HI disk (first dotted ellipse). The R-band image on the right 
shows a high surface brightness disk component that extends beyond R$_{25}$ (middle 
dashed-dotted ellipse) followed by a low surface brightness component that begins slightly within the 
break radius (outer-most dashed ellipse). The HI disk (inner-most dotted ellipse) lies well 
inside the high surface brightness disk component.

\section{GAS KINEMATICS \label{kin}}
In this section, we return to the HI synthesis observations to explore the neutral 
gas kinematics. In addition, we use integral field unit (IFU) spectroscopy to extract
ionized gas kinematics in the central region of NGC5005. We use both ionized and neutral gas kinematics to 
derive the circular rotational velocities of the gas as it follows the overall gravitational 
potential of the galaxy. Starting with the velocity field, we follow well established techniques 
to determine the bulk motions of the gas from which we extracted a rotation curve. This process is 
described in more detail below.
 
\subsection{Neutral Gas Kinematics \label{hikin}}
Beginning with the HI, we used standard analysis tools distributed as part of the GIPSY software 
package (\citealt{gipsy}) for analysis of HI synthesis data cubes to extract an intensity-weighted 
mean velocity field for NGC5005. To create this first moment map, individual channels from the low 
resolution datacube at 5 km s$^{-1}$ channel$^{-1}$ were smoothed by a factor of 3, and clipped at the 
3$\sigma$ level, then interactively blotted to identify signal. Figure \ref{n5005} (d) shows the final 
velocity field with isovelocity contours. The straight isovelocity contours seen in Figure \ref{n5005}
(d) indicate that the rotational velocity of the gas remains fairly constant with radius with no obvious
signs of a warp at large radii. Kinematic warps due to changes in position and/or inclination angle of 
the disk often result in a velocity field with a characteristic S-shaped distortion seen at larger radii 
(e.g., M83; \citealt{velfields}). The low signal-to-noise of the HI observations within 
the central $\sim$20\arcsec~(1.6 kpc at 16.5 Mpc) makes it difficult to trace the HI gas motions. However, 
as introduced previously, \cite{COdynamics} found an S-shaped distortion in the CO velocity field. This 
central kinematic distortion is most likely due to the presence of a bar causing the gas to be in 
non-circular orbits.

The filled circles in the bottom panel of Figure \ref{rotcur} show the rotation curve for NGC5005 
derived by fitting a series of concentric tilted rings to the HI velocity field (GIPSY task {\sc rotcur}). 
In this analysis the 
galaxy is described using a set of concentric rings, each with its own center coordinates 
($x_{\rm 0}$, $y_{\rm 0}$), systemic velocity ($V_{\rm sys})$, expansion velocity, inclination angle ({\it i}), 
position angle (P.A.), and rotation velocity ($V_{\rm rot}$). The radii and widths of the rings are
defined such that the rotation curve is sampled at a rate of two points per synthesized
beam width, which in this case is every 10\arcsec. The expansion velocity is fixed to zero.
Initial estimates for the center 
coordinates, inclination angle, and position angle were first obtained by fitting ellipses to the 
HI intensity distribution out to 3$\sigma$ (GIPSY task {\sc ellfit}). The systemic velocity was measured
from the spectral profile derived from the integrated intensity HI map. These initial values were 
fed into {\sc rotcur} and iteratively fit to the receding and approaching sides together.
Once the center position ($x_0$, $y_0$) was fixed, {\sc rotcur} was run again to determine
$V_{\rm sys}$ with {\it i} and P.A. as free parameters. Additional runs of {\sc rotcur} were
used with ($x_0$, $y_0$) and $V_{\rm sys}$ fixed to determine {\it i} and P.A. for a combination
of both sides. The rings are allowed to vary in position and inclination angle as a function of radius 
to account for warps ({\it top panels}). However, point-to-point variations in {\it i} can be caused by 
effects such as streaming motions along spiral arms for which the fitting routine tries to compensate by
changing {\it i} (\citealt{dmthings}).
A smoothly varying or constant distribution in {\it i} and P.A. was adopted to derive the underlying bulk 
rotation and prevent spurious second-order effects in the rotation curve. Because variations in position 
and inclination angle were small, constant position and inclination angles were used for this analysis 
({\it dashed lines}). Smoothly varying fits to both the inclination 
and position angle  were also tried and determined to be unnecessary as this only changed the derived 
circular rotational velocities by a few km s$^{-1}$. A final run of {\sc rotcur} was done with all 
parameters fixed to derive the final azimuthally averaged rotation curve ({\it bottom panel, filled circles}).

Uncertainties on the rotational velocities were
conservatively estimated to be a combination of non-circular thermal gas motions and
inclination errors. The uncertainties on the rotational velocities for the first two radii are 
dominated by thermal gas motions and beam smearing. A velocity dispersion, or second
moment map, was created from the high resolution HI data cube. Uncertainties due to thermal gas
motions were estimated to be about 10\%~of the average velocity dispersion values, which 
is the accuracy to which we believe the line centroid can be estimated. The velocity dispersion for
NGC5005 in the radial range 20\arcsec~to 40\arcsec~is quite large; average values range between 40-50 km s$^{-1}$.
This is why the first two rotation curve points have noticeably larger error bars than the rotational
velocities farther out where the dispersion drops to more typical values on the order of 15 km s$^{-1}$.
Outside of r $\ga$ 40\arcsec, the uncertainties
are dominated by inclination angle errors. Due to the relatively high inclination angle ($\sim$68\degr) 
of NGC5005, an inclination angle error of $\pm$10\degr~in the inner parts and $\pm$5\degr~in the 
outer parts produces corresponding rotational velocity errors of $\la$10\%~and $\lesssim$5\%, respectively.
Circular rotational velocities and their errors in the central 20\arcsec~were not derived from the 
neutral gas data due to the low signal-to-noise in the HI data.

The rotation curve for NGC5005 is remarkably flat; it varies by less than 10 km s$^{-1}$ over a range of 
radii from 20\arcsec~to 140\arcsec~(1.6 kpc to 11.2 kpc at 16.5 Mpc). The last measured rotational
velocity is 265.2$\pm$22.0 km s$^{-1}$ which yields a dynamical mass of (1.83$\pm$0.30)$\times$10$^{11}$
M$_{\sun}$ at 140\arcsec~(11.2 kpc at 16.5 Mpc). If we assume that the rotation curve remains flat and
extrapolate out to the break radius, this results in a dynamical mass of (2.82$\pm$0.47)$\times$10$^{11}$ 
M$_{\sun}$ measured at 216.3\arcsec~(17.3 kpc at 16.5 Mpc).

The HI rotation curve for NGC5005 was additionally derived using a code for modeling asymmetries in
disk galaxies ({\sc DiskFit}: \citealt{diskfit}). {\sc DiskFit} uses a physically motivated model
rather than a parameterization of concentric rings. It is capable of fitting models with both
rotation and lopsided or bisymmetric non-circular motions, as would be expected due to the
presence of a bar (\citealt{DFex}). Rotation curves for NGC5005 were derived using three
different models: rotation only, lopsided flow, and bisymmetric flow. Due to the low signal-to-noise
ratio of the HI at radii smaller than 20\arcsec, only the rotation dominated parts of the velocity field
were fit. Rotation curves derived from all three {\sc DiskFit} models, therefore, were in good
agreement with a rotation only model and the tilted ring fits from {\sc rotcur}. Derived best
fitting kinematic parameters from GIPSY's {\sc rotcur} and {\sc DiskFit}'s rotation only model
are presented in Table \ref{kinprops}.

\begin{figure}[ht]
\plotone{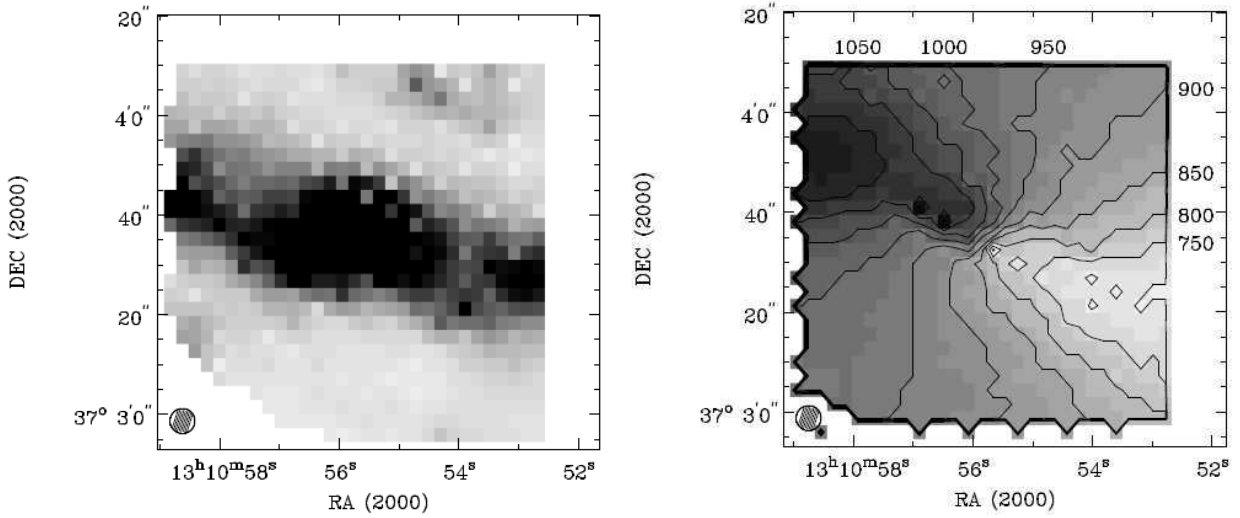}
\caption{{\it Left}: SparsePak [NII] $\lambda$6584 emission line flux map for the 
central 70\arcsec~$\times$ 70\arcsec~of NGC5005. The [NII] $\lambda$6584 flux map illustrates where 
the ionized gas emission is strongest and thus where the cross-correlation analysis has the highest
signal-to-noise ratio.
{\it Right}: Luminosity weighted mean velocity field for the central 70\arcsec~$\times$ 70\arcsec~from
SparsePak observations of ionized gas. Contours represent lines of constant rotational
velocity, the values of which are indicated above and on the left in units of km s$^{-1}$.
\label{spakvel}}
\end{figure}

\subsection{Ionized Gas Kinematics \label{spakkin}}
\begin{figure}
\plotone{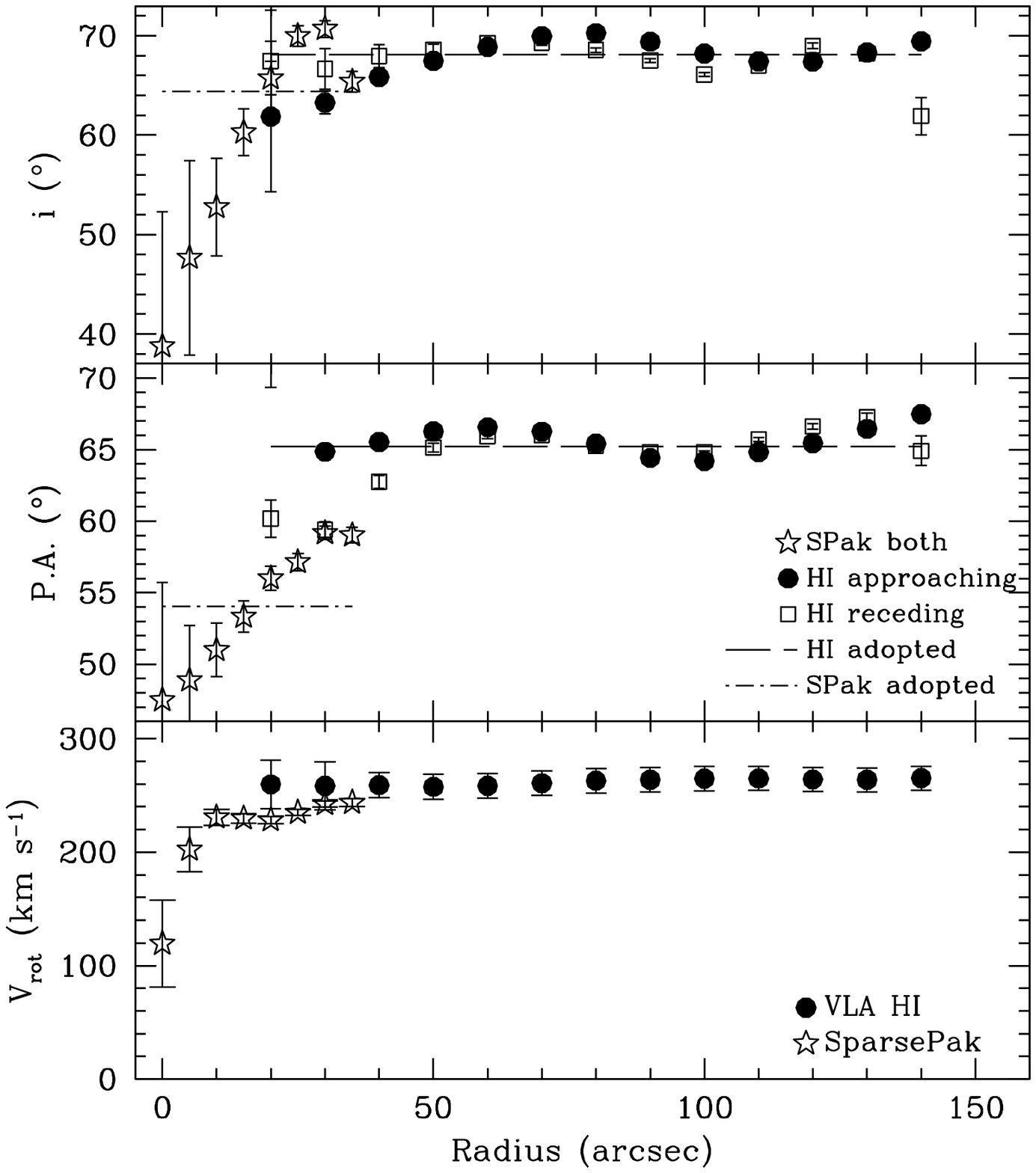}
\caption{Rotation curves for VLA HI and SparsePak (SPak) ionized gas derived from tilted rings using 
the GIPSY task {\sc rotcur}. 
The radial range fit for the HI was from 20\arcsec-140\arcsec~(1.6-11.2 kpc at 16.5 Mpc) and 
0\arcsec-35\arcsec~(0-2.8 kpc at 16.5 Mpc) for the ionized gas.
{\it Top}: Inclination angle fits to the approaching ({\it filled circles}) and receding
({\it open squares}) sides and the adopted constant inclination angle fit to both sides ({\it dashed line})
of the HI velocity field. Both sides of the SparsePak velocity field were fit at the same time
({\it open stars}), and a constant inclination angle was also adopted ({\it dot-dash line}).
{\it Middle}: Position angle fits to the approaching ({\it filled circles}) and receding
({\it open squares}) sides and the adopted constant position angle fit to both sides ({\it dashed line})
of the HI velocity field. The {\it open stars} show the SparsePak position angles fit to both sides
of the velocity field and {\it dash-dot line} shows the adopted constant position angle.
{\it Bottom}: Circular rotational velocities fit to both sides for the HI ({\it filled circles}) and 
SparsePak ionized gas ({\it open stars}) derived using the adopted inclination and position angles at each radius. 
\label{rotcur}}
\end{figure}

Given the low signal-to-noise ratio of the HI observations in the inner 20\arcsec~of NGC5005, we turn
to ionized gas to provide kinematic constraints in this inner region. Figure \ref{spakvel}
shows the ionized gas emission map as traced by the [NII]$\lambda$6584 line flux ({\it left}) and luminosity 
weighted mean velocity field determined from cross-correlation of the entire spectrum ({\it right}). We used 
the [NII]$\lambda$6584 emission line to demonstrate the distribution of ionized gas in
the central 70\arcsec~of NGC5005, as the H$\alpha$ flux is often affected by underlying stellar absorption
in the diffuse gas regions. Regions where the [NII]$\lambda$6584 line flux is strongest
represent areas with higher signal-to-noise ratios with typical TDR values around 40 from the 
cross-correlation results. The lower signal-to-noise ratio areas still have reliable cross-correlation results
with TDR values closer to 20.
We used {\sc rotcur} in GIPSY to derive a rotation curve from the ionized gas velocity field 
following the same steps for the HI data outlined above. The rotation curve derived from SparsePak ionized gas
observations is shown in Figure \ref{rotcur} ({\it open stars}). The ionized gas velocity field was sampled 
every 5\arcsec~out to a radius of 35\arcsec. We allowed the position angle, inclination angle, and the 
systemic velocity to be fit by {\sc rotcur} and not constrained to the values used for the HI derived 
rotation curve. Constant values of position and inclination angle were adopted ({\it dash-dot lines}). 
The ionized gas data fills in the central kinematics that are unconstrained by the HI observations and 
extends up towards the HI derived rotation velocity at 40\arcsec. The inner two HI rotation curve data points 
display higher rotational velocities than the ionized gas, but this is most likely due to the effects
of beam smearing.

While the ionized gas morphology agrees well with the CO morphology found by \cite{COdynamics}, we 
do not observe as large of a velocity in the nuclear disk (r $<$ 10\arcsec). This difference could
arise because the ionized gas rotational velocities were derived from the centroid of the emission line 
detected in a 5\arcsec~fiber and, therefore, do not include the high velocity dispersion of gas associated with the
AGN. We do, however, detect the same central S-shaped distortion in the ionized gas velocity field, 
as was detected by \cite{COdynamics} in the CO. Given the similar morphologies, it is plausible that 
the ionized gas is subjected to the same bar-driven dynamics as the molecular gas discussed
extensively in \cite{COdynamics}.

\begin{deluxetable}{lcc}
\tabletypesize{\footnotesize}
\tablecaption{Kinematic Properties of NGC5005 \label{kinprops}}
\tablewidth{0pt}
\tablehead{\colhead{Parameter} & \colhead{{\sc rotcur} Value} & \colhead{{\sc DiskFit} Value}}
\startdata
Kinematic center (RA, Dec J2000) & 13:10:56.28, +37:03:31.7 & 13:10:39.05, +37:05:35.4 \\
Systemic velocity & 948 km s$^{-1}$ & 944 km s$^{-1}$ \\
Adopted position angle & 65.6\arcdeg & 65.2\arcdeg \\
Adopted inclination angle & 68.1\arcdeg & 67.8\arcdeg \\
\enddata
\end{deluxetable}

\section{DARK MATTER MASS DECOMPOSITION \label{dmdecomp}}
Rotation curve decomposition analysis is a powerful tool to probe the dark matter content in
galaxies (e.g., \citealt{dmthings}). This well established analysis method combines the
derived galaxy kinematics from the observed rotation curve with estimates of the baryonic mass
distribution in order to derive an estimate of the dark matter distribution as a function of
radius (e.g., \citealt{rcdecomp}). \cite{dmthings} found that both widely used dark
matter models (NFW and isothermal) fit equally well for galaxies with M$_B$ $<$ -19. 
In this analysis, we fit only the spherical pseudo-isothermal 
dark matter halo model to the observations as the data are not sensitive to the central dark
matter distribution. 

One of the strengths of the present analysis is the use of NIR data
to derive stellar bulge and disk distributions. In particular, the {\it Spitzer} 3.6$\mu$m
images are relatively insensitive to radial changes in mass-to-light (M/L) ratio that might
be introduced by either internal dust obscuration or changes in the dominant stellar population.
Indeed, the M/L ratio at 3.6$\mu$m is nearly independent of star formation history 
(e.g., \citealt{nirml}, \citealt{3.6ML}). The combination of NIR, HI, and CO observations allow many of the 
uncertainties associated with the spatial distribution of both major baryonic components 
(gas and stars) to be minimized. 

\begin{figure}
\plotone{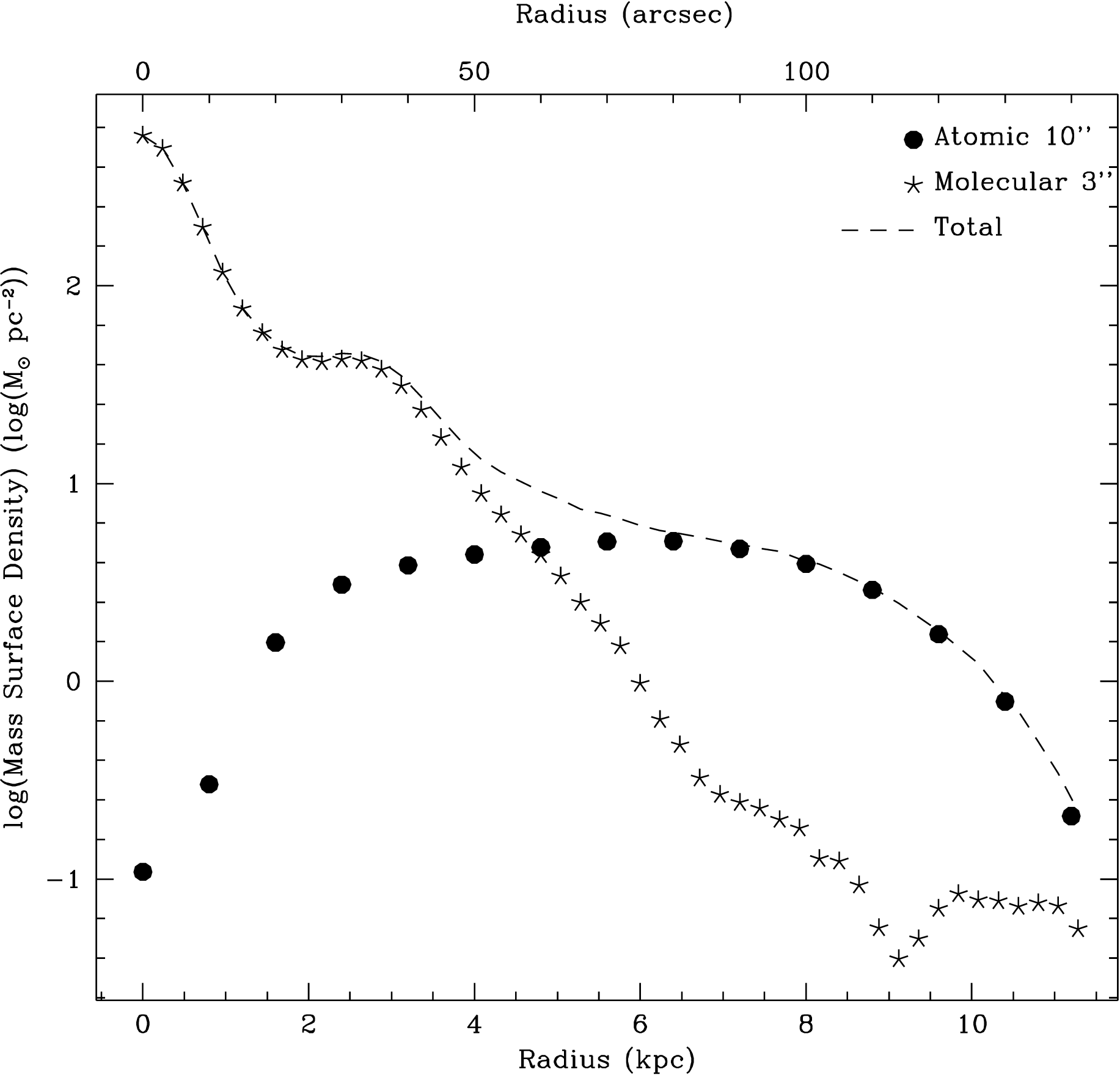}
\caption{Atomic and molecular gas mass surface density distributions for NGC5005. The atomic 
({\it filled circles}) and molecular ({\it stars}) gas distributions are derived from the HI and CO 
integrated intensity maps, respectively (see Figure \ref{n5005} (a)). The spatial sampling for each distribution 
is roughly half the beam size from their respective maps (20\arcsec~for HI and 6\arcsec~for CO). The 
dashed line represents the total gas mass surface density when the atomic and molecular distributions are added 
together. Mass surface densities are plotted as the log of the value.
\label{gasmsd}}
\end{figure}

The contributions from gas and stars to the overall rotation curve for NGC5005 were calculated 
from the gas mass and stellar light surface density profiles. The gas mass surface density was
measured from the HI and CO integrated intensity maps. The gas distributions were derived
separately and then added together to create a total gas mass surface density distribution. The
contribution to the total circular rotational velocity was calculated from the total gas distribution 
using the GIPSY task {\sc rotmod} assuming an infinitely thin gas disk. Following 
\cite{Broeils}, the total gas mass was multiplied by a factor 1.3 to account for primordial helium.
Figure \ref{gasmsd} shows the gas mass surface density distributions for the HI and
H$_2$ (as traced by CO)
separately ({\it filled circles} and {\it stars}, respectively), as well as the resulting total gas 
distribution when added together ({\it dashed line}). The molecular gas is the dominant contribution
to the total gas mass surface density for radii within 60\arcsec~(4.8 kpc at 16.5 Mpc). The bump
in the molecular gas distribution between 30\arcsec-40\arcsec~corresponds to the ring of emission observed in
the narrowband H$\alpha$ image (Section \ref{optdata} and Figure \ref{ellipses}).
Model rotation curves for the stellar distributions were derived separately for the bulge and disk 
based on the decomposed light surface density profiles discussed in Section \ref{3.6data}. The
contribution to the total circular rotational velocity for these stellar components were calculated
with {\sc rotmod}. A vertical sech-squared distribution with a vertical scale height 
of 0.93 kpc estimated using the fiducial relation derived
in \cite{DMII} was used for the stellar disk model. A spherical distribution was assumed for the bulge 
model rotational velocities. 

\begin{figure}
\plotone{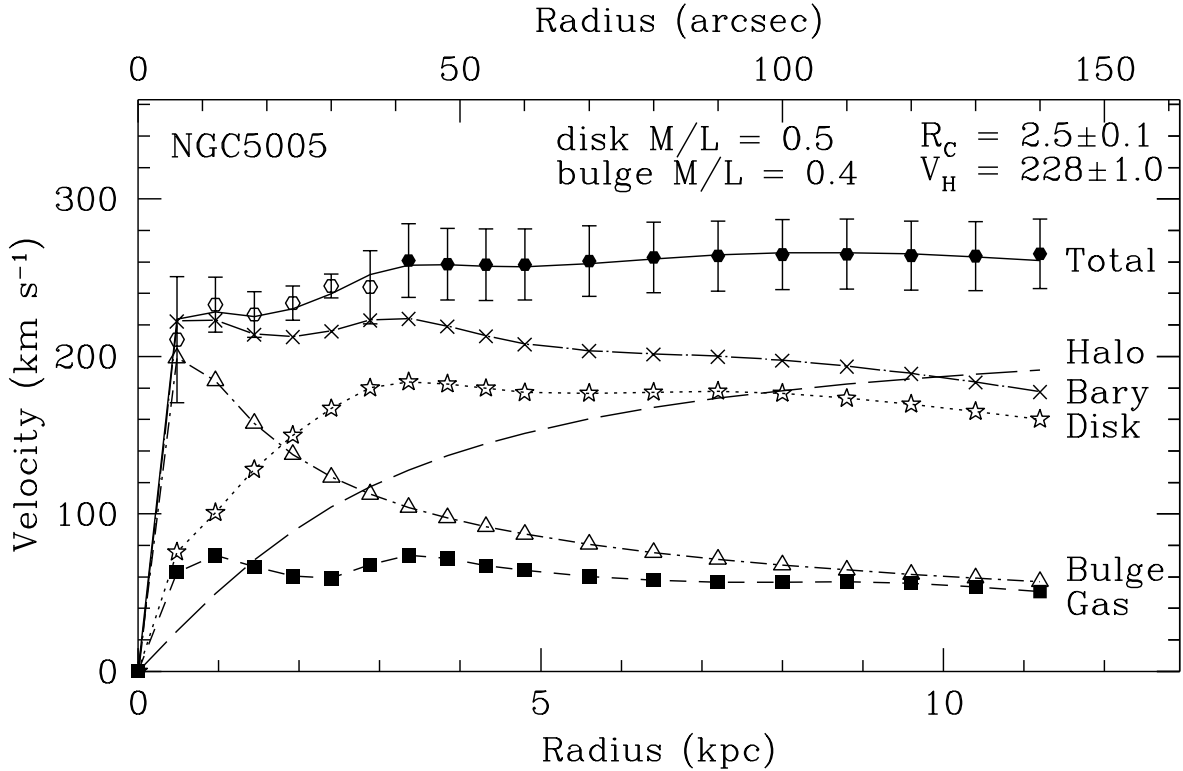}
\caption{Decomposition of the rotation curve from the HI, optical, and NIR data. The total gas 
({\it filled squares}), stellar bulge ({\it open triangles}), stellar disk ({\it open stars}), and 
dark matter halo model ({\it dashed line}) components are added in quadrature to achieve the best
overall fit ({\it solid line}) to the observed rotation curve. Open circles represent circular
rotation velocities derived from the ionized gas (SparsePak) observations and the filled circles
are from the HI (VLA) observations. Adding the baryonic components in quadrature
without the dark matter results in the crosses. This total baryonic contribution primarily follows 
the stellar distribution in this gas-deficient galaxy.
\label{rotfit}}
\end{figure}

Figure \ref{rotfit} presents the rotation curve decomposition for NGC5005. Circular rotation velocities 
for the total gas, stellar bulge, and stellar disk were summed in quadrature and subtracted from the 
observed rotation curve. A spherical pseudo-isothermal dark matter halo model was fit to the residuals. 
The bulge and disk M/L ratios were left as free parameters to be fit during the decomposition. They were 
assumed to be constant with radius. For this galaxy, leaving the stellar disk and bulge M/L ratios in 
addition to the characteristic radius (R$_{\rm C}$) and velocity (V$_{\rm H}$) of the dark matter halo 
model as free parameters creates a poorly constrained problem without a unique solution. Therefore, the 
best fitting R$_{\rm C}$, V$_{\rm H}$, M/L$_{\rm bulge}$, and M/L$_{\rm disk}$ were achieved using interactive 
fitting. The best fit achieved in this analysis resulted in a disk M/L$_{\rm disk}$ = 0.5 
and a bulge M/L$_{\rm bulge}$ = 0.4. This disk M/L is consistent with \cite{LMCML}
who suggest that the M/L ratio at 3.6$\mu$m should be about 0.5 based on stellar population models and
measurements of the Large Magellanic Cloud. The bulge and disk M/L ratios result in a best-fitting dark matter
halo model with a characteristic radius of 2.5$\pm$0.1 kpc and a velocity of 228$\pm$1.0 km s$^{-1}$.
The observed rotation curve could not be adequately described at large radii using the baryonic 
contributions alone (crosses in Figure \ref{rotfit}).

The dip in velocity in the observed ionized gas rotation curve between 12\arcsec-24\arcsec~(0.96-
1.92 kpc at 16.5 Mpc) is an interesting feature which demonstrates the connection between the galaxy
morphology and its kinematics. It appears to occur in the transition region between the nuclear disk 
and ring structure, as seen in the H$\alpha$ image (Figure \ref{ellipses} {\it left}) and detected in CO 
emission (\citealt{COdynamics}). This region corresponds to a decrease in the total gas contribution 
to the overall rotation curve, which is dominated by the molecular gas at these inner
radii (Figure \ref{gasmsd}), as well as to where the stellar 
contribution to the overall rotation
curve transitions from being bulge- to disk-dominated. It is also visible as a prominent dip in the
equivalent width profile seen in Figure \ref{phot} (d).

The mass decomposition reveals that NGC5005 is baryon dominated at the inner-most radii. NGC5005
follows the same trend as most other massive, luminous galaxies where its mass is baryon dominated
in the core and gradually becomes more dark matter dominated with increasing radius (e.g., 
\citealt{dmthings}). Furthermore, galaxies of similar dynamical mass from \cite{dmthings}, 
such as NGC6946 and NGC7331, display rotation curve decompositions similar to that of NGC5005
where the dark matter halo contribution to the total rotation curve does not overtake the
stellar disk contribution to the total rotation curve until close to or beyond the last measured point.
The fractional contribution of the maximum
stellar disk to the rotation curve at 2.2 times the disk scale length is only 0.71 in NGC5005.
This implies that the stellar disk can only explain about 70\%~of the observed rotation curve
at the point where the stellar disk's contribution should be greatest.
This is slightly less than average values found for other samples of galaxies (\citealt{dmdwarfs} 
and references therein), which were between 0.8 and 0.9. 

\section{CONCLUSION \label{conclusion}}
We have acquired multiwavelength observations to examine the gas and stellar content in NGC5005.
We have utilized new VLA C configuration observations to explore the neutral hydrogen gas
morphology and kinematics, as well as SparsePak IFU observations to derive kinematics of
the ionized gas in the central region of NGC5005. Deep 3.6$\mu$m images from {\it Spitzer} allow
us to trace the extended stellar population. Finally, we have used optical broadband B and R 
and narrowband H$\alpha$ observations to detect properties of the dominant stellar population and 
star formation activity. The main results from this multiwavelength study are summarized below.
\begin{enumerate}
\item The most striking morphological feature of NGC5005 is its extended stellar population. The 
neutral and molecular gas components are both confined well within the high surface brightness 
stellar disk. The low surface brightness stellar component appears to extend nearly 3 times beyond
the gas disk detected in HI. The central 40\arcsec~of NGC5005 has a lack of detectable HI, but 
displays both molecular and ionized gas emission.
\item Spectra of the ionized gas in the central region of NGC5005 reveal signatures of both thermal
and non-thermal emission processes. The nuclear AGN region displays broad emission lines with non-thermal
line intensity ratios. Double-peaked emission lines are seen to the northwest of the nucleus, approximately
where \cite{COdynamics} found dynamical evidence for infalling gas potentially due to an inconspicuous 
stellar bar. The ionized gas emission outside of the nuclear disk appears to be dominated by thermal
processes from recent star formation activity.
\item Inspection of the optical broadband images reveals numerous known and suspected 
companions of NGC5005. These include SDSS J131051.05+365623.4 which was detected in the VLA
observations and SDSS J131105.57+371036.1 which appears to be possibly connected to NGC5005
by a faint stellar stream.
\item Surface photometry of NGC5005 shows a NIR surface brightness distribution with a
central bulge component and an inner and outer exponential disk. The radial profile
of optical color displays a distribution that is typical for large spirals like NGC5005,
and is consistent with NGC5005 having little current star formation activity. The unusually
low star formation rate and equivalent width measurements from the H$\alpha$ imaging are
expected given the gas poor nature of NGC5005.
\item A rotation curve for NGC5005 was derived from both the neutral and ionized gas kinematics.
The ionized gas rotation curve in the central region of NGC5005 exhibits structure which can be
connected directly to the ionized gas morphology (which follows the molecular gas), and also
appears to correspond to the stellar distribution as it transitions from bulge to disk dominated.
The HI rotation curve at large radii is remarkably flat and exhibits no signs of warps.
\item The gas kinematics and surface densities were used in conjunction with the stellar bulge and disk
light profiles to decompose the observed rotation curve of NGC5005 and derive a model dark matter
distribution. The results of the mass decomposition demonstrate that NGC5005 is baryon dominated
in its central region and that the prominent stellar disk only maximally contributes about
70\%~to the total rotation curve.
\end{enumerate}

\acknowledgments
The authors acknowledge observational and technical support from the National 
Radio Astronomy Observatory (NRAO). We acknowledge use of the WIYN
0.9m telescope operated by WIYN Inc. on behalf of a Consortium of ten partner Universities 
and Organizations, and the WIYN 3.5m telescope.
WIYN is a joint partnership of the University of Wisconsin at Madison, Indiana University, 
Yale University, and the National Optical Astronomical Observatory. This work is based 
on observations made with the Spitzer Space Telescope, which is operated by the Jet Propulsion 
Laboratory, California Institute of Technology under a contract with NASA. Emily E. Richards
acknowledges support from the Provost's Travel Award for Women in Science, a professional development 
fund supported through the Office of the Provost at Indiana University Bloomington. This research has 
made use of the NASA/IPAC Extragalactic Database (NED) which is operated by the Jet Propulsion Laboratory, 
California Institute of Technology, under contract with the National Aeronautics and Space Administration.

\end{document}